\documentclass[12pt]{article}

\usepackage{amssymb}
\usepackage{amsmath}

\usepackage{graphics}
\usepackage{epsfig}

\begin{document}

\title{\bf Hadron multiplicities in $\mathbf{e^+e^-}$ annihilation with
heavy primary quarks}

\author{A.V. Kisselev\thanks{E-mail: alexandre.kisselev@ihep.ru} \
and V.A. Petrov\thanks{E-mail: vladimir.petrov@ihep.ru} \\
\small Institute for High Energy Physics, 142281 Protvino, Russia}

\date{}

\maketitle

\thispagestyle{empty}

\bigskip

\begin{abstract}
The multiple hadron production in the events induced by the heavy
primary quarks in $e^+e^-$ annihilation is reconsidered with
account of corrected experimental data. New value for the
multiplicity in $b\bar{b}$ events is presented on the basis of
pQCD estimates.
\end{abstract}

%%%%%%%%%%%%%
% Main text %
%%%%%%%%%%%%%

\section{Introduction}

The so-called ``na\"{\i}ve model''~\cite{Rowson:85, Kisselev:88}
was the first attempt to give a framework for calculating the
multiplicity of hadrons produced in addition to decay products of
the heavy quark-antiquark pair in $e^+e^-$ annihilation. Later on,
it was argued~\cite{Schumm:92} that the difference between
multiplicities in heavy and light quark events ($l = u, \, d, \,
s$),
\begin{equation}\label{delta_Ql}
\delta_{Ql} = N_{Q\bar{Q}}(W) - N_{l\bar{l}}(W),
\end{equation}
tends to a constant value at high collision energy:
\begin{equation}\label{MLLA_prediction}
\delta_{Ql} \rightarrow \delta_{Ql}^{MLLA} = 2n_Q -
N_{l\bar{l}}(m_Q^2 \mathbf{e}).
\end{equation}
Here and in what follows, $N_{Q\bar{Q}}$ and $N_{l\bar{l}}$ are
mean multiplicities of charged hadrons in heavy and light quark
events, respectively.%
\footnote{Everywhere below, it is assumed that we deal with
\emph{mean} multiplicities of \emph{charged} hadrons.}

The comparison with the data has shown that the ``na\"{\i}ve
model'' describes the data on $\delta_{bl}$ up to $W = 58$
GeV~\cite{Rowson:85, b_events_low_energies_1,
b_c_events_low_energies, b_events_low_energies_2}, but
underestimates the LEP and SLAC
data~\cite{b_events_high_energies_1, b_c_events_high_energies,
b_events_high_energies_2}. As for the so-called MLLA
formula~\eqref{MLLA_prediction}, it significantly overestimates
both low and high-energy data on $\delta_{bl}$.

\begin{figure}[ht]
\begin{center}
\epsfysize=11cm \epsffile{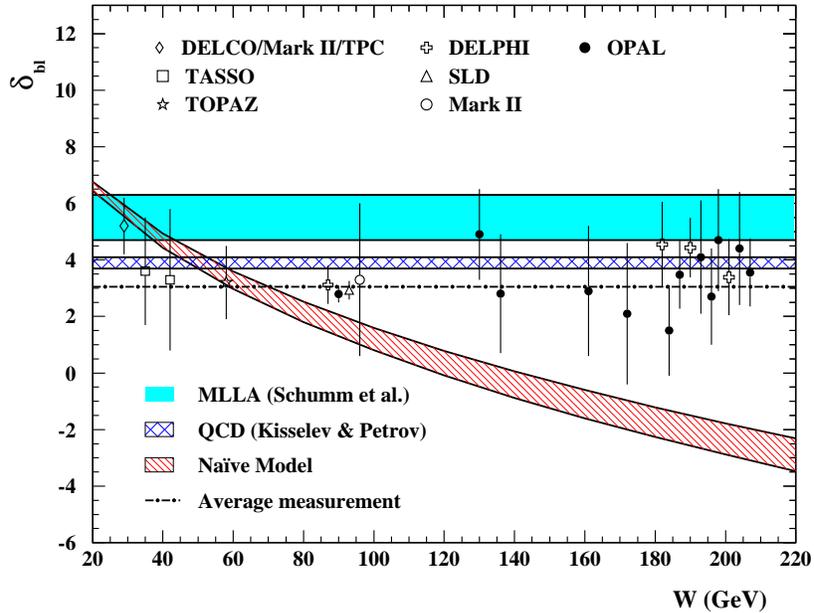}
\end{center}
\caption{QCD prediction~\cite{Petrov:95} and MLLA
result~\cite{Schumm:92} vs. experimental data on the multiplicity
difference $\delta_{bl}$. The data are not corrected as in
Ref.~\cite{Dokshitzer:06}. The prediction of the ``na\"{\i}ve
model'' is also shown.} \label{fig:data_delta_bl}
\end{figure}

The detailed QCD calculations of the difference between associated
multiplicities of charged hadron in $e^+e^-$ annihilation were
made in \cite{Petrov:95}. The QCD expressions for $\delta_{Ql}$
from Ref.~\cite{Petrov:95} appeared to be in a good agreement with
experimental measurements of associated hadron multiplicities in
$e^+e^-$ annihilation (see, for instance, \cite{Chrin:94,
Metzger:95}). Note that up to now, our formula provided \emph{the
best description} of all the available data on $\delta_{bl}$, see
Fig.~\ref{fig:data_delta_bl}.

Moreover, we made a prediction for $\delta_{cl}$~\cite{Petrov:95}.
It is also in a very good agreement with all the data on
$\delta_{cl}$~\cite{Rowson:85, b_c_events_low_energies,
b_c_events_high_energies}. Let us stress that the very value of
$\delta_{cl}$ was derived in \cite{Petrov:95} \emph{before the
precise measurements of $\delta_{cl}$} were
made~\cite{b_c_events_high_energies}, that allows to
test QCD calculations.%
\footnote{In the low energy measurements~\cite{Rowson:85,
b_c_events_low_energies}, the total error of $\delta_{cl}$ was
about $\pm 1.5$.}

As we will see below, it is the hadron multiplicity in the light
quark events that enables one to calculate the multiplicity
differences $\delta_{Ql}$. The mean charged multiplicities in $l
\bar{l}$ events at different energies corrected for detector
effects as well as for initial state radiation were recently cited
in \cite{Dokshitzer:06}. The corrected multiplicity differences
averaged over all presently published results were also
presented~\cite{Dokshitzer:06}:
\begin{eqnarray}
\delta_{bl}^{\rm exp} &=& 3.12 \pm 0.14 \;, \label{delta_bl_exp}
\\
\delta_{cl}^{\rm exp} &=& 1.0 \pm 0.4 \;. \label{delta_cl_exp}
\end{eqnarray}

The first goal of this paper is to re-estimate our QCD predictions
for the quantity $\delta_{bl}$, taking into account the corrected
experimental data on $N_{l\bar{l}}(W)$ from \cite{Dokshitzer:06}.
The second goal is to argue that the MLLA
formula~\eqref{MLLA_prediction} is nothing but some part of our
QCD expression (see Section~\ref{sec:I}), and, as the comparison
to the data shows, it should be regarded as a rather rough
approximation of the QCD result.%
\footnote{The shortcomings of the MLLA formula were already
briefly discussed in Ref.~\cite{Petrov:95}.}

\section{QCD formula for multiplicity difference}
\label{sec:I}

The hadron multiplicity in a $q\bar{q}$ event, $N_{q\bar{q}}(W)$,
looks like~\cite{Petrov:95}
\begin{equation}\label{mult_Y}
N_{q\bar{q}}(Y) = 2\,n_{q} + \int\limits_0^Y \! d \eta \,
\hat{n}_g(Y - \eta) \, E_q(\eta) \;,
\end{equation}
where variables
\begin{equation}\label{eta}
\eta = \ln \frac{W^2}{k^2}
\end{equation}
and
\begin{equation}\label{Y}
Y = \ln \frac{W^2}{Q_0^2}\;
\end{equation}
are introduced. In what follows, the notation $q=Q$ will mean
charm or beauty quarks, while the notation $q=l$ will correspond
to a massless case (when a pair of $u,\ d$ or $s$-quarks is
produced, whose masses are assumed to be zero).

The first term in the r.h.s. of Eq.~\eqref{mult_Y}, $2n_{q}$, is
the multiplicity of primary hadron decay products. It is extracted
from the data ($2n_c = 5.2$, $2n_b = 11.0$~\cite{Schumm:92}, and
$2n_l = 2.4$~\cite{Chrin:94}).

The term $E_q(k^2/W^2)$ in \eqref{mult_Y} is the inclusive
spectrum of a gluon jet with a virtuality up to $k^2$ emitted by
primary quarks.%
\footnote{It was explained in detail in Ref.~\cite{Petrov:95} that
one should not consider this mechanism of hadron production via
gluon jets as due to ``a single cascading gluon'', as some authors
believe~\cite{Dokshitzer:06}. That $E_q$ is an inclusive spectrum
of the gluon jets is seen, e.g., from the fact that $\int
(dk^2\!/k^2) \,E_q(k^2/W^2) > 1$.}
It is defined by the discontinuity of of the two-gluon irreducible
$\gamma^{*}g^{*}$ ($Z^{*}g^{*}$) amplitude normalized to the total
$e^+e^-$ rate. The quantity $\hat{n}_g(k^2)$ is related to
$n_g(k^2)$, the mean multiplicity of hadrons inside this jet:
\begin{equation}\label{n_g_reduced}
\hat{n}_g (k^2) = \frac{C_F \, \alpha_s(k^2)}{\pi} \, n_g(k^2) \;.
\end{equation}
Here $\alpha_s(k^2)$ is a strong coupling constant, and $C_F =
(N_c^2 - 1)/2 N_c$, with $N_c$ being the number of colors.

The physical meaning of the function
\begin{equation}\label{cent_mult_Y}
N_q(Y) = \int\limits_0^Y \! d \eta \, \hat{n}_g(Y - \eta) \,
E_q(\eta) \;
\end{equation}
in Eq.~\eqref{mult_Y} is the following. It describes the average
number of hadrons produced in virtual gluon jets emitted by
primary quark (antiquark) of the type $q$. In other words, it is
the multiplicity in $q\bar{q}$ event except for the multiplicity
of the decay products of these quarks at the final stage of
hadronization (the first term in \eqref{mult_Y}).

\begin{figure}[ht]
\epsfysize=6cm \epsffile{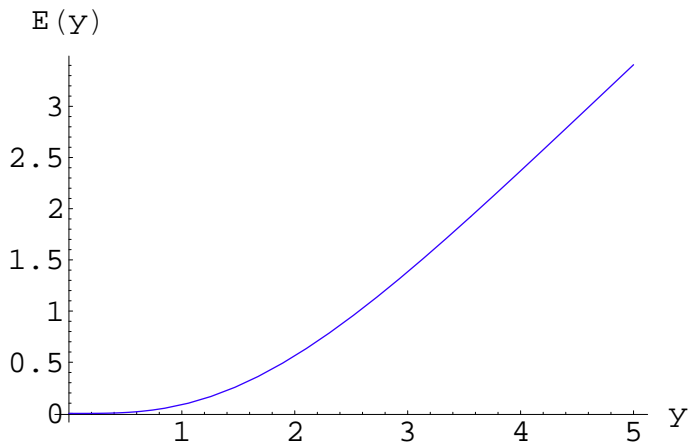} \caption{The function $E(y \equiv
\eta)$.}
\label{fig:E}
\end{figure}

For the massless case, the function $E \equiv E_l$ was calculated
in our paper~\cite{Petrov:95}. In terms of a dimensionless
variable
\begin{equation}\label{sigma_vs_eta}
\sigma = \exp (- \eta)
\end{equation}
it looks like
\begin{eqnarray}\label{E}
E(\eta(\sigma)) &=& (1 + 2\sigma + 2\sigma^2) \ln \frac{1}{\sigma}
- \frac{3 + 7\sigma}{2} (1 - \sigma) - \sigma (1 + \sigma) \left(
\ln \frac{1}{\sigma} \right)^2
\nonumber \\
&+& 4\sigma(1 + \sigma) I(\sigma) \;,
\end{eqnarray}
with
\begin{equation}\label{I}
I(\sigma) = \int\limits_{\sigma}^1 \! \frac{dx}{1 + x} \, \ln
\frac{1}{x} \equiv \frac{\pi^2}{4} - \mathrm{\rm Li}_2(1 + \sigma)
\;,
\end{equation}
where $\mathrm{\rm Li}_2(z)$ is the Euler dilogarithm. The
function $E(\eta)$ is presented in Fig.~\ref{fig:E}. It has the
asymptotics $E(\eta)|_{\eta \rightarrow \, \infty} =
E^{(asym)}(\eta) = \eta - 1/2$.

The derivative of $E(\eta)$ is positive, as one can see in the
next Fig.~\ref{fig:der_E}, with $\partial E(\eta)/\partial \eta =
0$ at $\eta=0$, and $\partial E(\eta)/\partial \eta = 1$ at
$\eta=\infty$. As a result, associative multiplicity $N_q(W)$
\eqref{cent_mult_Y} is a monotone increasing function of the
energy $W$ for any positive function $n_g(k^2)$.%
\footnote{It results from the relation $\partial N_q(Y)/\partial Y
= \int_0^Y \! d \eta \, \hat{n}_g(\eta) \, \partial E(Y -
\eta)/\partial Y$.}
\begin{figure}[ht]
\epsfysize=6cm \epsffile{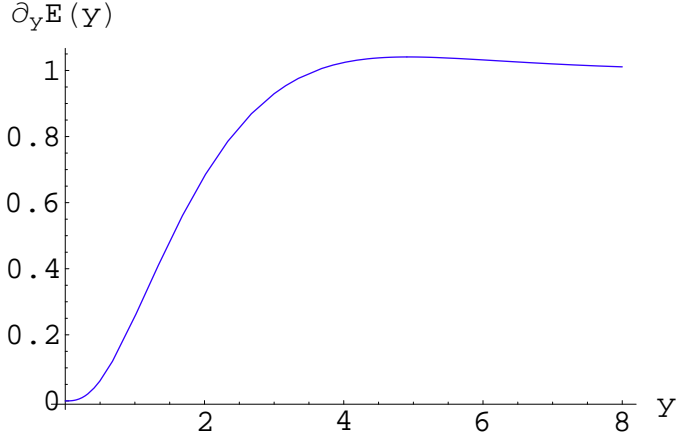} \caption{The function
$\partial_y E(y \equiv \eta)$.}
\label{fig:der_E}
\end{figure}

Now let us consider the difference between multiplicities in heavy
and light quark events, $\delta_{Ql}$, which is defined by
Eq.~\eqref{delta_Ql}. The following representation was found in
Ref.~\cite{Petrov:95}:
\begin{equation}\label{delta_Ql_QCD}
\delta_{Ql}^{QCD} = 2(n_Q - n_l) - \Delta N_Q(Y_m) \;.
\end{equation}
Here new notation,
\begin{equation}\label{delta_N}
\Delta N_Q(Y_m) = N_q - N_Q = \int\limits_{-\infty}^{Y_m} \! dy \,
\hat{n}_g (Y_m - y) \, \Delta E_Q(y) \;,
\end{equation}
as well as variables
\begin{equation}\label{y}
y = \ln \frac{m_Q^2}{k^2}
\end{equation}
and
\begin{equation}\label{Ym}
Y_m = \ln \frac{m_Q^2}{Q_0^2} \;
\end{equation}
are introduced.

The non-trivial result which was obtained in Ref.~\cite{Petrov:95}
is that the function
\begin{equation}\label{delta_E_def}
\Delta E_Q = E_l - E_Q \equiv E - E_Q
\end{equation}
depends only on a variable
\begin{equation}\label{ro_vs_y}
\rho = \exp (-y) \;,
\end{equation}
but not on energy W. The explicit form of $\Delta E_Q$ is known to
be
\begin{eqnarray}\label{Delta_E}
\Delta E_Q(y(\rho)) &=& (1 - 3\rho + \frac{7}{2} \, \rho^2) \ln
\frac{1}{\rho} + \rho (7\rho - 20) \, J(\rho) + \frac{20}{\rho -
4} [1 - J(\rho)]
\nonumber \\
&+& 7\rho + \frac{9}{2} \;,
\end{eqnarray}
where
\begin{equation}\label{J}
J(\rho) =
  \begin{cases}
    \sqrt{\frac{\rho}{\rho - 4}}
    \ln \left( \frac{\sqrt{\rho} + \sqrt{\rho - 4}}{2}
    \right),
    & \rho > 4 \cr
    \ 1 \;, & \rho = 4 \cr
    \sqrt{\frac{\rho}{4 - \rho}}
    \arctan \left( \frac{\sqrt{4 - \rho}}{\rho} \right),
    & \rho < 4 \; .
  \end{cases}
\end{equation}

\begin{figure}[ht]
\epsfysize=6cm \epsffile{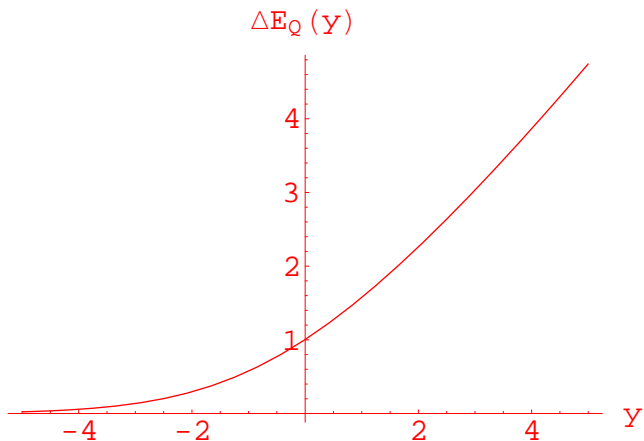}
\caption{The function $\Delta E_Q(y)$.}
\label{fig:deltaE}
\end{figure}

Since $\Delta E_Q(y)$ has the asymptotics
\begin{equation}\label{Delta_E_asym}
\Delta E_Q(y) \Big|_{y \rightarrow -\infty} \simeq \frac{11}{3} \,
\exp(-|y|) \;,
\end{equation}
the integral in Eq.~\eqref{delta_N} converges rapidly at $y
\rightarrow -\infty$. The function $\Delta E_Q(y)$ is shown in
Fig.~\ref{fig:deltaE}. We find that $\Delta E_Q(y)|_{y \rightarrow
\, \infty} = \Delta E_Q^{(asym)}(y) = y - 3/2$.

One should mention the following important relation:%
\footnote{Here (and below) $\mathbf{e}$ means the base of the
natural logarithm.}
\begin{equation}\label{diff_E_asym_reg}
\Delta E_Q(y - 1) - E(y) \Big|_{y \rightarrow \, \infty}  \simeq
\frac{5}{2} \, \sqrt{\mathbf{e}} \, \ln 2 \, \exp(-y/2) \;.
\end{equation}
In other words,
\begin{equation}\label{diff_E_asym}
\Delta E_Q(y) \simeq E(y+1)
\end{equation}
\emph{at large} $y$.

If one puts $\Delta E_Q(y) = E(y+1)$, then (neglecting also the
contribution from the region $y < -1$):
\begin{equation}\label{delta_N_asym}
\Delta N_Q = N_{l\bar{l}}(m_Q^2 \mathrm{e}) - 2\,n_l \;.
\end{equation}
Correspondingly, the approximate expression for $\delta_{Ql}$ is
then of the form:
\begin{equation}\label{delta_asym}
\delta_{Ql}^{(appr)} = 2n_Q - N_{l\bar{l}}(m_Q^2 \mathrm{e}) =
\delta_{Ql}^{MLLA} \;,
\end{equation}
where $\delta_{Ql}^{MLLA}$ is the MLLA prediction for the
multiplicity difference from Ref~\cite{Schumm:92}. Remember that
the function $N_{l\bar{l}}(W)$ describes the hadron multiplicity
in light quark event at colliding energy $W$.

\begin{figure}[ht]
\epsfysize=6cm \epsffile{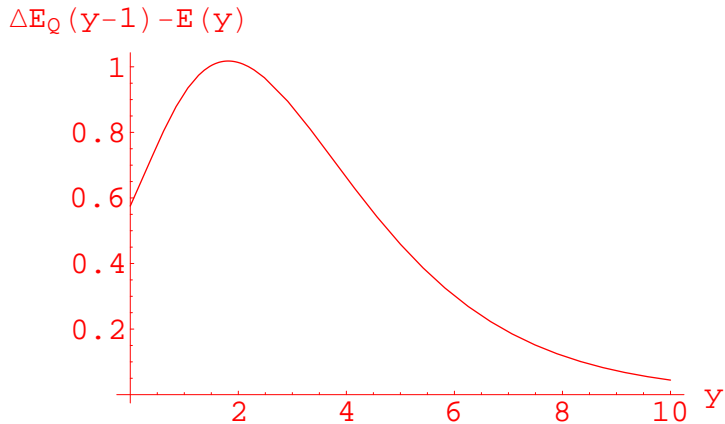} \caption{The
difference $\Delta E_Q(y - 1) - E(y)$ as a function of the
variable $y$.}
\label{fig:deltaE_vs_E}
\end{figure}

However, Eq.~\eqref{diff_E_asym} is very far from being satisfied
at relevant $y < Y_m$,%
\footnote{For the beauty case, one has $Y_m \lesssim 3.2$.}
as it is clearly seen in Fig.~\ref{fig:deltaE_vs_E}. As a result,
there could be a large difference between
$\delta_{Ql}^{(appr)}$~\eqref{delta_asym} and QCD expression
$\delta_{Ql}^{QCD}$~\eqref{delta_Ql_QCD}.

To demonstrate this, it is convenient to represent expression for
$\Delta N_Q$~\eqref{delta_N} in the form:
\begin{eqnarray}\label{Delta_N_expan}
\Delta N_Q(Y_m) &=& \int\limits_{0}^{Y_m + 1} \! dy \, \hat{n}_g
(Y_m + 1 - y) E(y)
\nonumber \\
 &+&  \int\limits_{-\infty}^{-1} \! dy \, \hat{n}_g (Y_m - y)
\, \Delta E_Q(y)
\nonumber \\
&+&  \int\limits_{0}^{Y_m + 1} \! dy \, \hat{n}_g (Y_m + 1 - y)
[\Delta E_Q(y - 1) - E(y)]
\nonumber \\
&\equiv& [N_{l\bar{l}}(m_Q^2 \mathrm{e}) - 2\,n_l ] + \delta
N_Q^{(1)}(Y_m ) + \delta N_Q^{(2)}(Y_m) \;,
\end{eqnarray}
that results in  the formula (see Eq.~\eqref{delta_Ql_QCD})
\begin{eqnarray}\label{delta_asym_corr}
\delta_{Ql}^{QCD} &=& 2n_Q - N_{l\bar{l}}(m_Q^2 \mathrm{e})-
\delta N_Q^{(1)}(Y_m) - \delta N_Q^{(2)}(Y_m)
\nonumber \\
&=& \delta_{Ql}^{(appr)} - \delta N_Q^{(1)}(Y_m ) - \delta
N_Q^{(2)}(Y_m)\;.
\end{eqnarray}
Here we have introduced the notations
\begin{equation}\label{delta_N1_def}
N_Q^{(1)}(Y_m) = \int\limits_{-\infty}^{-1} \! dy \, \hat{n}_g
(Y_m - y) \, \Delta E_Q(y) \;,
\end{equation}
and
\begin{equation}\label{delta_N2_def}
N_Q^{(2)}(Y_m) = \int\limits_{0}^{Y_m + 1} \! dy \, \hat{n}_g (Y_m
+ 1 - y) [\Delta E_Q(y - 1) - E(y)] \;.
\end{equation}
Note, both $N_Q^{(2)}(Y_m)$ and $N_Q^{(2)}(Y_m)$ are positive
functions, since $\Delta E_Q(y) > 0$ at all $y$ and $\Delta E_Q(y
- 1) - E(y) > 0$ at $y \geqslant 0$ (see Fig.~3 and Fig.~4).

In order to exploit the corrected data on $N_{l\bar{l}}(W)$ at $W
=8 \mathrm{\ GeV}$,
\begin{equation}\label{N(mbe)_MLLA}
N_{l\bar{l}}(8.0 \mathrm{\ GeV}) = 6.70 \pm 0.34 \;,
\end{equation}
we have chosen the mass of b-quark to be $m_b = 4.85$ GeV, which
corresponds to $m_b \sqrt{\mathbf{e}} = 8$ GeV.

The estimates show that the dominant correction to
$\delta_{Ql}^{QCD}$ is $\delta N_Q^{(2)}$, not  $\delta
N_Q^{(1)}$. To calculate a lower bound of $\delta N_b^{(2)}$, let
us use the following inequality:
\begin{equation}\label{deltaE_E}
\Delta E_Q(y) = E(y + \Delta y_Q) \;.
\end{equation}
Note that $\Delta y_Q$ is a monotone non-increasing function of $y
\geqslant 0$ and it tends to 1 at large $y$. It solves the
equation:
\begin{equation}\label{delta_y}
\Delta E_Q(Y_m) \geqslant E(Y_m + \Delta y_Q) \;,
\end{equation}
where $Y_m$ is defined above~\eqref{Ym}. Then we get from
Eqs.~\eqref{delta_N2_def} and \eqref{deltaE_E}:
\begin{equation}\label{delta_N2_bound}
\delta N_Q^{(2)} \geqslant N_{l\bar{l}}(Y_m + \Delta y_Q) -
N_{l\bar{l}}(Y_m + 1) - \int\limits_{0}^{\Delta y_Q - 1} \! dy \,
\hat{n}_g (Y_m + \Delta y_Q - y) E(y) \;.
\end{equation}

For our further estimates, we need to know the hadron multiplicity
in light quark events in the energy interval $2.5 \mathrm{\ GeV}
\leqslant W \leqslant 28 \mathrm{\ GeV}$. By fitting the data on
hadron multiplicity in the light quark events at low $W$, we get
the expression:
\begin{equation}\label{N_light_fit}
N_{l\bar{l}}(W) = 2.07 + 1.11 \  \ln W + 0.54 \ \ln^2 W \;.
\end{equation}

Putting $Q_0 = 1$ GeV, we find $\Delta y_b = 1.61$. Taking into
account that the last term in
Eq.~\eqref{delta_N2_bound} is negligible,%
\footnote{Since $E(y) < 0.02$ in the region $0 \leqslant y
\leqslant \Delta y_b -1 = 0.61$.}
we get from \eqref{delta_N2_bound}, \eqref{N_light_fit}:
\begin{equation}\label{delta_N2_b_num}
\delta N_b^{(2)} \geqslant 1.07\;.
\end{equation}
Correspondingly, our prediction accounting the revision of the
data on the multiplicity in the light quark events,
\begin{equation}\label{delta_bl_upper_bound}
\delta_{bl}^{QCD} \leqslant 2 n_b - N_{l\bar{l}}(Y_m + \Delta y_b)
= 3.33 \pm 0.38 \;,
\end{equation}
appears to be lower than our previous result $\delta_{bl} =
3.68$~\cite{Petrov:95}. We used the value
\begin{equation}\label{2n_b}
2n_b = 11.10 \pm 0.18 \;.
\end{equation}
The error of $N_{l\bar{l}}$ was taken to be $\pm \, 0.34$. Let us
stress that our upper bound \eqref{delta_bl_upper_bound} is very
close to the present experimental value of
$\delta_{bl}^{exp}$~\eqref{delta_bl_exp}.

Now let us derive a lower bound on $\delta_{bl}^{QCD}$. To do
this, let us start from Eq.~\eqref{delta_N}. It is convenient to
represent the integral in \eqref{delta_N} as a sum of two terms:%
\footnote{We took into account that the region $-\infty < y < -4$
gives a negligible contribution to $\Delta N_b$.}
\begin{eqnarray}\label{delta_N_bl}
\Delta N_b &=& \int\limits_{-4}^{-1} \! dy \, \hat{n}_g (Y_b - y)
\, \Delta E_b(y) + \int\limits_{-1}^{Y_b} \! dy \, \hat{n}_g (Y_m
- y) \, \Delta E_b(y)
\nonumber \\
&=& \Delta N_b^{(1)} + \Delta N_b^{(2)} \;,
\end{eqnarray}
with $Y_b = \ln (m_b^2/Q_0^2) \simeq 3.16$. Consider the first
term in \eqref{delta_N_bl}. One can check that
\begin{equation}\label{Delta_vs_E_1}
\Delta E(y) < 0.18 \, E(y + 5.8)
\end{equation}
in the region $-4 < y < -1$, that leads to the inequality
\begin{equation}\label{Delta_Nb_1}
\Delta N_b^{(1)} < 0.18 \int\limits_{1.8}^{4.8} \! dy \, \hat{n}_g
(Y_b + 5.8 - y) \, \Delta E_b(y) \;.
\end{equation}

The estimations show that $\hat{n}_g (Y_b + 5.8 - y) < 2 \,
\hat{n}_g (4.8 - y)$ when $y$ varies from 1.8 to 4.8. Thus, we
get:
\begin{equation}\label{Delta_Nb_1_estimate}
\Delta N_b^{(1)} < 0.36 \, \left[ N_{l\bar{l}}(W=11 \mathrm{\
GeV}) - N_{l\bar{l}}(W=2.5 \mathrm{\ GeV}) \right] = 1.54 \pm 0.17
\;.
\end{equation}

The second term in \eqref{delta_N_bl} can be estimated by using
the inequality
\begin{equation}\label{Delta_vs_E_2}
\Delta E(y) < 0.62 \, E(y + 3.5)
\end{equation}
which is valid in the region  $-1 < y < Y_b$. Then
\begin{equation}\label{Delta_Nb_2_estimate}
\Delta N_b^{(2)} < 0.62 \left[ N_{l\bar{l}}(W=28 \mathrm{\ GeV}) -
N_{l\bar{l}}(W=3.5 \mathrm{\ GeV}) \right] = 4.61 \pm 0.30 \;.
\end{equation}

\begin{figure}[ht]
\begin{center}
\epsfysize=7cm \epsffile{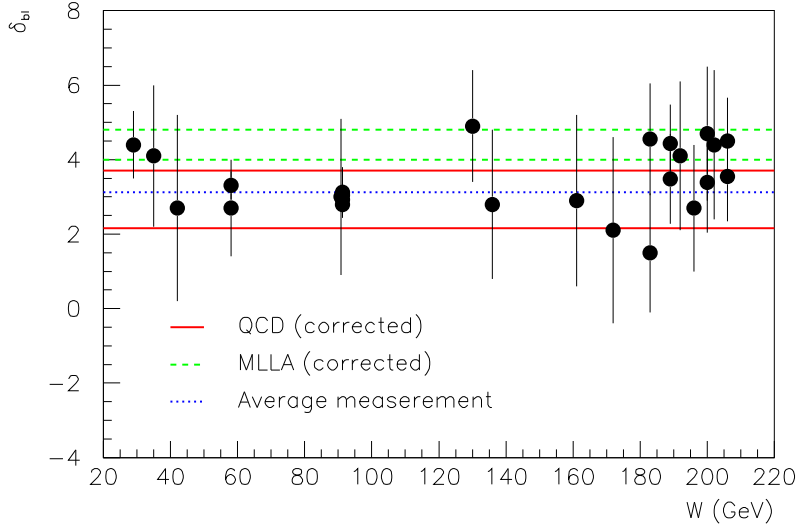}
\end{center}
\caption{Our new QCD result for $\delta_{bl}$ (a corridor between
the two solid lines) as well as MLLA prediction from
Ref.~\cite{Dokshitzer:06} (a corridor between the two dashed
lines) vs. \emph{corrected} experimental data.}
\label{fig:data_corr_delta_bl}
\end{figure}

As a result, we obtain from Eqs.~\eqref{delta_Ql_QCD},
\eqref{delta_N} and \eqref{Delta_Nb_1_estimate},
\eqref{Delta_Nb_2_estimate} the lower bound on
$\delta_{bl}^{QCD}$:
\begin{equation}\label{delta_bl_lower_bound}
\delta_{bl}^{QCD} > 2.55 \pm 0.39 \;.
\end{equation}
Fig.~\ref{fig:data_corr_delta_bl} demonstrates that our QCD
predictions for $\delta_{bl}$ are very close to the corrected
experimental data.

Our results can be compared with the MLLA expectation reported
recently in Ref.~\cite{Dokshitzer:06}:
\begin{equation}\label{bl_MLLA}
\delta_{bl}^{MLLA} = 4.4  \pm 0.4 \;.
\end{equation}
Note that the scheme of Ref.~\cite{Dokshitzer:06} is not stable
against next-to-MLLA corrections. According to Eq.~A(30) from
\cite{Dokshitzer:06}, the MLLA prediction \eqref{delta_asym} is
modified as follows:
\begin{equation}\label{Ql_NMLLA}
\delta_{Ql}^{NMLLA} = 2n_Q - N_{l\bar{l}}(m_Q^2 \mathrm{e}) \Big\{
1 + \frac{3 \alpha_s(m_Q)}{2 \pi} \Big[ \frac{\pi^2}{24} + \Big(
\frac{\pi^2}{3}  - \frac{5}{4} \Big)\Big]\Big\}\;.
\end{equation}
The next-to-MLLA corrections in \eqref{Ql_NMLLA} change the
result~\eqref{bl_MLLA} to
\begin{equation}\label{bl_NMLLA}
\delta_{bl}^{NMLLA} = 2.6 \pm 0.4 \;.
\end{equation}
The situation is worse in the case of c-quark. The
formula~\eqref{Ql_NMLLA} results in a unsatisfactory low value
\begin{equation}\label{cl_NMLLA}
\delta_{cl}^{NMLLA} = -0.1 \pm 0.4 \;.
\end{equation}
This demonstrates us once more that the lowest-order MLLA
expression~\eqref{MLLA_prediction} is not correct.

Moreover, as we have shown above, the deviation of the function
$\Delta E_Q(y)$ from $\Delta E_Q^{(asym)}(y) = y - 3/2$, as well
as the deviation of $E(y)$ from $E^{(asym)}(y) = y - 1/2$, cannot
be neglected. In other words, the MLLA formula
\eqref{delta_asym} is, in fact, not a full QCD result.%
\footnote{It explains why formula \eqref{bl_MLLA} overestimates
the data by more than one unit.}
It is nothing but \emph{a part} of the correct QCD formulae
\eqref{delta_Ql}, \eqref{delta_N} in a very rough approximation
$E(y) = \Delta E_Q(y -1) $. So it is senseless to try to
``improve'' it with next-to-MLLA calculations.

\section{Conclusions}

We have derived the QCD formula for the difference between hadron
multiplicities in heavy and light quark events in $e^+e^-$
annihilation (with $Q$ being a type of a heavy quark):
\begin{eqnarray}\label{delta_bQ_concl}
\delta_{Ql}^{QCD} &=& 2n_Q - N_{l\bar{l}}(m_Q^2 \mathrm{e})
\\ \nonumber
&-& \int\limits_{Q_0^2}^{m_Q^2 \mathrm{e}} \! \frac{d k^2}{k^2} \,
\hat{n}_g (k^2) \left[ \Delta E_Q \left( \frac{m_Q^2}{k^2} \right)
- E \left( \frac{m_Q^2 \mathrm{e}}{k^2} \right) \right]
\\ \nonumber
&-& \int\limits_{m_Q^2 \mathrm{e}}^{\infty} \! \frac{d k^2}{k^2}
\, \hat{n}_g (k^2) \, \Delta E_Q \left( \frac{m_Q^2}{k^2} \right)
\;.
\end{eqnarray}
Here $n_g(k^2)$ describes the mean number of charged hadrons in
the gluon jet with the virtuality up to $k^2$, and $E$, $\Delta
E_Q$ are known functions.

By using the data on the hadron multiplicity in light quark events
$N_{l\bar{l}}$, corrected for the detector effects and initial
state radiation effects~\cite{Dokshitzer:06}, we have obtained
from \eqref{delta_bQ_concl} the bounds:
\begin{equation}\label{delta_bl_QCD}
2.2 < \delta_{bl}^{QCD} < 3.7 \;.
\end{equation}
Let us note that this estimate does not depend on a specific
choice of the function $n_g(k^2)$, and it is in a good agreement
with the average experimental value $\delta_{bl}^{\rm exp} = 3.12
\pm 0.14$.

Two last terms in \eqref{delta_bQ_concl} are positive and
numerically large.%
\footnote{In particular, the second term in \eqref{delta_bQ_concl}
(dominating the third one) is equal to 1.1 for the case of the
beauty pair production ($m_Q = m_b$, $n_Q = n_b$).}
As a result, a deviation of the MLLA prediction,
\begin{equation}\label{delta_b_MLLA_concl}
\delta_{bl}^{MLLA} = 2n_b - N_{l\bar{l}}(m_b^2 \mathrm{e}) \;,
\end{equation}
from the QCD expression,%
\footnote{This formula is an equivalent compact form of
Eq.~\eqref{delta_bQ_concl} for $Q=b$.}
\begin{equation}\label{delta_bl_concl}
\delta_{bl}^{QCD} = 2(n_b - n_l) - \int\limits_{Q_0^2}^{m_b^2} \!
\frac{d k^2}{k^2} \, \hat{n}_g (k^2) \, \Delta E_Q \left(
\frac{m_b^2}{k^2} \right) \;,
\end{equation}
appears to be significant.

As one can see, the MLLA formula is a too rough approximation of
the QCD formula. The former results from the latter on the
assumption that the quantities $E(y)$ and $\Delta E_Q(y)$ can be
replaced by their asymptotics at $y \rightarrow \infty$. Since the
relevant values of $y$ are far from being very large, this
assumption is not correct, and it leads to a significant
overestimation of $\delta_{bl}$. Thus, any attempt to use the MLLA
expression \eqref{MLLA_prediction} as a first-order approximation
for higher-order calculations (as it is done in
\cite{Dokshitzer:06} ) is poorly justified.

%%%%%%%%%%%%%%%%%%%%
% Acknowledgements %
%%%%%%%%%%%%%%%%%%%%

\section*{Acknowledgements}

We thank professor V.~Khoze for illuminating conversations.

%%%%%%%%%%%%%%
% References %
%%%%%%%%%%%%%%


\begin{thebibliography}{99}

\bibitem{Rowson:85}
MARK II Collaboration: P.C.~Rowson \emph{et al.}, Rev. Lett. {\bf
54} (1985) 2580;
\bibitem{Kisselev:88}
A.V.~Kisselev, V.A.~Petrov and O.P. Yuschenko, Z. Phys. C {\bf 41}
(1988) 521.
\bibitem{Schumm:92}
B.A.~Schumm, Yu.L.~Dokshitzer, V.A.~Khoze and D.S.~Koetke, Phys.
Rev. Lett. {\bf 69} (1992) 3025.
\bibitem{b_events_low_energies_1}
DELCO Collaboration: M.~Sakuda \emph{et al.}, Phys. Lett. B {\bf
161} (1985) 412.
\bibitem{b_c_events_low_energies}
TPC Collaboration: H.~Aihara \emph{et al.}, Phys. Lett. B {\bf
184} (1987) 299; TASSO Collaboration: W.~Brauschweig, \emph{et
al.},  Z. Phys. C {\bf 42} (1989) 17; TASSO Collaboration:
M.~Althoff \emph{et al.}, Phys. Lett. B {\bf 42} (1984) 243.
\bibitem{b_events_low_energies_2}
TOPAZ Collaboration: K.~Nagai \emph{et al.}, Phys. Lett. B {\bf
278} (1992) 506; VENUS Collaboration: K.~Okabe \emph{et al.},
Phys. Lett. B {\bf 423} (1998) 407.
\bibitem{b_events_high_energies_1}
MARK II Collaboration: B.A.~Schumm \emph{et al.}, Phys. Rev. D
{\bf 646} (1992) 453; DELPHI Collaboration: P.~Abreu \emph{et
al.}, Phys. Lett. B {\bf 347} (1995) 447.
\bibitem{b_c_events_high_energies}
OPAL Collaboration: R.~Akers \emph{et al.}, Phys. Lett. B {\bf
352} (1995) 176; SLD Collaboration: K.~Abe \emph{et al.}, Phys.
Lett. B {\bf 386} (1996) 475; SLD Collaboration: K.~Abe \emph{et
al.}, Phys. Rev. D {\bf 69} (2004) 072003.
\bibitem{b_events_high_energies_2}
DELPHI Collaboration: P.~Abreu \emph{et al.}, Phys. Lett. B {\bf
479} (2000) 118; Erratum, \emph{ibid} B {\bf 492} (2000) 398; OPAL
Collaboration: G.~Abbiendi \emph{et al.}, Phys. Lett. B {\bf 550}
(2002) 33.
\bibitem{Petrov:95}
V.A.~Petrov and A.V.~Kisselev, Z. Phys. C {\bf 66} (1995) 453;
Nucl. Phys. (Proc. Suppl.) B {\bf 39}, C (1995) 364.
\bibitem{Chrin:94}
DELPHI Collaboration: J. Chrin \emph{et al.}, in \emph{Proc. of
the 27-th International Conference on High Energy Physics},
Glasgow, UK, 20-27 July 1994, eds. P.J.~Bussey and I.G.~Knowles,
p.~893.
\bibitem{Metzger:95}
W.~Metzger, in \emph{Proc. of the International Europhysics
Conference on High Energy Physics}, Brussel, Belgium, 27 July- 2
August 1995, eds. J.~Lemonne, C.~Vander Velde and F.~Verbeure
(World Sientific, Singapore), p.~323.
\bibitem{Dokshitzer:06}
Yu.L.~Dokshitzer, F.~Fabbri, V.A.~Khoze and W.~Ochs, Eur. Phys. J.
C {\bf 45} (2006) 387.

\end{thebibliography}
\end{document}